# LeqMod: Adaptable Lesion-Quantification-Consistent Modulation for Deep Learning Low-Count PET Image Denoising

Menghua Xia, *Member, IEEE*, Huidong Xie, *Student Member, IEEE*, Qiong Liu, *Student Member, IEEE*, Bo Zhou, *Member, IEEE*, Hanzhong Wang, Biao Li, Axel Rominger, Quanzheng Li, *Senior Member, IEEE*, Ramsey D. Badawi, *Senior Member, IEEE*, Kuangyu Shi, Georges El Fakhri, *Fellow, IEEE*, and Chi Liu, *Senior Member, IEEE*

*Abstract*—Deep learning-based positron emission tomography (PET) image denoising offers the potential to reduce radiation exposure and scanning time by transforming low-count images into high-count equivalents. However, existing methods typically blur crucial details, leading to inaccurate lesion quantification. This paper proposes a lesion-perceived and quantification-consistent modulation (LeqMod) strategy for enhanced PET image denoising, via employing downstream lesion quantification analysis as auxiliary tools. The LeqMod is a plug-and-play design adaptable to a wide range of model architectures, modulating the sampling and optimization procedures of model training without adding any computational burden to the inference phase. Specifically, the LeqMod consists of two components, the lesion-perceived modulation (LeMod) and the multiscale quantification-consistent modulation (QuMod). The LeMod enhances lesion contrast and visibility by allocating higher sampling weights and stricter loss criteria to lesion-present samples determined by an auxiliary segmentation network than lesion-absent ones. The QuMod further emphasizes quantification accuracy for both the mean and maximum standardized uptake value ($SUV_{mean}$ and $SUV_{max}$) across multiscale sub-regions throughout the entire image, thereby reducing biases of denoised results relative to high-count references. Experiments conducted on large PET datasets from multiple centers and vendors, and varying noise levels demonstrated the LeqMod efficacy across various denoising frameworks. Compared to frameworks without LeqMod, the integration of LeqMod reduces the lesion $SUV_{max}$ bias by 5.92% on average and increases the peak signal-to-noise ratio (PSNR) by 0.36 on average, when denoising images across participating sites. (Code is available at https://github.com/mhxiaaa/LeqMod_PET_denoising)

## I. INTRODUCTION

Positron emission tomography (PET) is a highly sensitive nuclear medicine imaging technique that can visualize functional processes in the body through injected radiotracers [1]. Its application spans a wide range of medical fields, including oncology, neurology, and cardiology, playing a crucial role in disease diagnosis and follow-up therapy [2]. The noise level in PET images is primarily determined by counts of detected photon events, which are influenced by three main factors: the injected radiotracer activity, the scanning time, and the efficiency of scanner hardware. Reducing the tracer dose for enhanced safety, shortening the scanning time for a better patient experience and fewer motion artifacts, and cutting hardware costs all result in low-count images with high noise. To address the tradeoff between high-quality imaging and constraints imposed by clinical and economic considerations, software denoising algorithms are developed to remove noise and enhance the quality of low-count images.

Published PET image denoising algorithms encompass both conventional and deep learning (DL) approaches [3]. Conventional methods consist of dedicatedly designed filters, such as the non-local mean (NLM) filter with spatiotemporal enhancements [4], the multiple-reconstruction NLM (MR-NLM) filter that leverages the redundant information from multiple PET reconstructions [5], wavelet transforms enhanced by anatomical knowledge [6], multiresolution curvelet transforms [7], and integrated wavelets and curvelets [8]. In recent years, DL denoising methods for PET images have thrived and outperformed conventional ones [9]. These DL models typically utilize paired low-count and high-count PET images as training datasets to learn a mapping function that translates low-count images into their high-count equivalents. These methods vary in their adopted network architectures, including convolutional neural networks (CNNs) [10][11][12],

This work was supported by the National Institutes of Health (NIH) under grant R01CA275188. Corresponding author: Chi Liu (email: chi.liu@yale.edu).
M. Xia, G. El Fakhri, and C. Liu are with the Department of Radiology and Biomedical Imaging, Yale University School of Medicine, USA.
H. Xie, Q. Liu, B. Zhou, and C. Liu are with the Department of Biomedical Engineering, Yale University, USA.
H. Wang and B. Li are with the Department of Nuclear Medicine, Ruijin Hospital, Shanghai Jiao Tong University School of Medicine, China.
A. Rominger and K. Shi are with the Department of Nuclear Medicine, Bern University Hospital, University of Bern, Switzerland.
K. Shi is also with the Computer Aided Medical Procedures and Augmented Reality, Institute of Informatics, Technical University of Munich, Germany.
R. D. Badawi is with the Department of Radiology, University of California Davis Medical Center, USA.
Q. Li is with the Department of Radiology, Massachusetts General Hospital and Harvard Medical School, USA.

generative adversarial networks (GANs) [13], and vision transformers (ViTs) [14]. Additionally, several models arm the denoising framework with added functionalities, such as embedding noise-aware mechanisms to adaptively handle varying low-count noise levels [15][16], deploying federated deep transformation networks to address privacy issues across multiple institutions [17], and utilizing magnetic resonance-guided fusion modules to exploit cross-modality information thereby boosting performance [18][19]. The latest advances involve integrating deep networks with traditional iterative optimization models [20] and exploring diffusion probabilistic models for PET denoising [21][22][23], both showing promising future potential.

Although DL methods for PET image denoising achieve visually appealing results, they tend to over-smooth images with diminished structural details. When applied to Fluorine-18 Fluorodeoxyglucose ($^{18}$F-FDG) PET images in oncology, which is the focus of this work, over-smoothed DL-denoised images could be insufficient for lesion detection and accurate radioactivity quantification. Typically, lesions appear with reduced contrast in denoised images or may even be erroneously omitted [24][25], causing underestimations in clinical metrics such as the standard uptake value (SUV) and false negatives in cancer diagnosis. Occasionally, DL models could be misled by significant noise, generating hallucinations of false-positive lesions [24][25]. This issue of lesion quantification inaccuracy in PET image denoising hinders the clinical applications of DL technologies.

To address this challenge, it is intuitive to consider employing either manually annotated lesion labels or automated lesion segmentation to facilitate the denoising process. Yet, to the best of our knowledge, no DL denoising method that incorporates this concept has been proposed for PET imaging. This oversight could be attributed to the significant difficulty in labeling lesion regions or acquiring datasets that include matched lesion labels for low- and high-count images. While some approaches for other medical imaging modalities have explored combining denoising with segmentation, they struggle to adapt to PET scenarios. For instance, networks integrating prior anatomical knowledge have been developed for low-dose CT denoising [26][27][28], where the anatomical information is derived from a subnetwork tasked with organ segmentation and characterization. Similarly, the DenoiSeg model [29], tailored for microscopy image analysis, leverages a single Unet with four output channels to concurrently perform both self-supervised denoising and supervised segmentation. Its multi-channel output delivers both denoised images and segmentation masks. These segmentation-assisted denoising approaches fit in situations where target objects occupy a relatively large area in images. They are not well-suited for PET cases where the targets, i.e., metabolically active lesions, may be extremely small (e.g., less than 150 voxels in an image of 440×440×644 voxels, or 674 mm³ out of 5.6×10⁸ mm³). Additionally, methods [26][27][28] rely on cascaded attention modules to fuse features from segmentation branches into the denoising network, complicating the primary denoising architecture and consequently increasing computational burden and inference time. The DenoiSeg model [29], while simple, is primarily focused on segmentation, treating denoising as a secondary task without an emphasis on denoising efficacy. Moreover, all these approaches use a denoising-segmentation-coupled design, which necessitates matched manual segmentation labels for noisy and clean image pairs in denoising datasets. The difficult access to these kinds of datasets might be the reason hindering the emergence of similar methods in PET.

In this study, to address the challenge of retaining lesion visibility and quantification accuracy in $^{18}$F-FDG PET cancer image denoising and to facilitate smooth deployment on DL models, we introduce the lesion-perceived and quantification-consistent modulation (LeqMod) strategy. Major contributions are as follows:

(1) The LeqMod utilizes auxiliary lesion segmentation and quantification to aid DL-based PET image denoising. It avoids the need for manual lesion labels in denoising datasets, by decoupling the auxiliary analysis from the main denoising process and pre-training the auxiliary tool on public sources. This design makes the LeqMod a versatile plug-and-play module compatible with a wide range of DL frameworks. Additionally, the LeqMod operates only during model training, adjusting sampling and optimization procedures without adding computational burden to model inference.

(2) The LeqMod introduces two novelties: lesion-perceived modulation (LeMod) and multiscale quantification-consistent modulation (QuMod). The LeMod leverages the pre-trained lesion segmentation network to enhance lesion visibility and lesion quantification accuracies. The QuMod further demands precise radioactivity quantifications on multiscale sub-regions across the entire image, thus improving overall image quality with mitigated over-smoothing.

(3) Extensive experiments were conducted on large datasets from multiple medical centers, major vendor scanners, and various low-count noise levels. First, the LeqMod effectiveness in enhancing lesion visibility and image quality was proven on Unet baseline. Then, detailed ablation studies illustrated individual contributions of each LeqMod component. Finally, the flexible applicability of the LeqMod across various denoising frameworks was demonstrated.

## II. METHODS

Fig. 1 illustrates the proposed method. The LeqMod operates solely during the training phase and is omitted in the inference. In terms of the denoising baseline, various network architectures can be selected. As a large-size 3D PET image cannot be processed as a whole due to computational limitations, it is systematically processed patch by patch. Let $I_{LC}$ / $I_{HC}$ denote the low/high-count PET image pairs. The denoising network $f_{den}(\cdot)$ is trained under supervisions of extracted low/high-count patch pairs $V_{LC}/V_{HC}$ and the guidance of the LeqMod. In the inference stage, the trained $f_{den}(\cdot)$ maps low-count patches $V_{LC}$ into denoised ones $V_{DEN}$, which are then reassembled into the denoised PET image $I_{DEN}$. Details are as below.

## A. Lesion-Perceived Modulation

The first component of the LeqMod, the LeMod, is designed to enhance the lesion contrast and visibility in denoised images. To accomplish this, the LeMod incorporates a segmentation network $f_{seg}(\cdot)$ to identify lesion regions.

Given that the lesion segmentation serves as an auxiliary task to support the denoising process without requiring perfect accuracy, the $f_{seg}(\cdot)$ within LeMod adopts a simple and compact Unet architecture [30]. This design comprises a four-layer hierarchical encoder and a three-layer decoder, with the shallowest layer's feature channels set to 32. To promote feature learning and facilitate gradient flow, residual connections are employed at each encoder layer, while deep supervisions are deployed at each decoder layer. Before being integrated into the denoising workflow, the $f_{seg}(\cdot)$ undergoes pre-training using public PET data sources (described in Section III-A) specifically curated for lesion segmentation tasks. Input the high-count patch $V_{HC}$, the trained $f_{seg}(\cdot)$ outputs the segmentation probability map $f_{seg}(V_{HC}(j)) \in [0,1]$, indicating the likelihood of each voxel $j$ belonging to a lesion.

The LeMod directly leverages lesion probability maps instead of their binarized masks, thus in a soft modulation way, to guide the denoising learning process to concentrate on lesions, through patch-level lesion-perceived sampling and voxel-level lesion-perceived consistency constraint.

### 1) Lesion-perceived sampling

After dividing PET images into patches, the vast majority of these patches do not contain lesions, as lesions are typically much smaller than the overall image size, as shown in the sample histogram of patch counts versus lesion probability in Fig. 1. The imbalanced distribution of patches with and without lesions will skew DL algorithms toward predicting the majority type of lesion-absent patches, detrimentally affecting the denoising performance on actual lesions. To this end, the LeMod designs a lesion-perceived sampling function to prioritize the minority yet crucial patches containing lesions. This function actively assigns greater sampling weights to lesion-present patches compared to lesion-absent ones, thereby mitigating the training data distribution imbalance between the lesion and non-lesion knowledge. The sampling weight for a training patch pair $(V_{LC}, V_{HC})$ is set as follows:

$$w(V_{LC}, V_{HC}) = \eta(V_{LC}) \cdot \max\left\{\max_j\{f_{seg}(V_{HC}(j))\}, w_{min}\right\} \quad (1)$$

where $f_{seg}(V_{HC}(j)) \in [0,1]$ is the lesion probability determined by the segmentation network on the high-count patch $V_{HC}$, signifying the likelihood of each voxel $j$ being part of a lesion. The highest lesion probability within a patch, denoted as $\max_j\{f_{seg}(V_{HC}(j))\}$, indicates the presence of lesions in $V_{HC}$. To prevent lesion-absent patches with near-zero probabilities from being overlooked during the weighted sampling process, while still prioritizing lesion-present patches, a minimum weight threshold $w_{min}$ is set (0.3 in this study). Additionally, considering the varying noise levels in low-count PET images and the increased challenge of recovering lesion details from noisier images, we further modulate sampling weights using a noise-aware factor $\eta(V_{LC})$. This factor is inversely related to the count level of $V_{LC}$, increasing as the count level decreases. Following the guidance in [31] for training a one-size-fits-all denoiser, which recommends allocating a greater proportion of samples to more challenging cases, the noise-aware factor $\eta(V_{LC})$ in this study is set to 0.35, 0.25, 0.15, 0.12, 0.08, and 0.05 for low-count levels of 1%, 2%, 5%, 10%, 25%, and 50%, respectively.

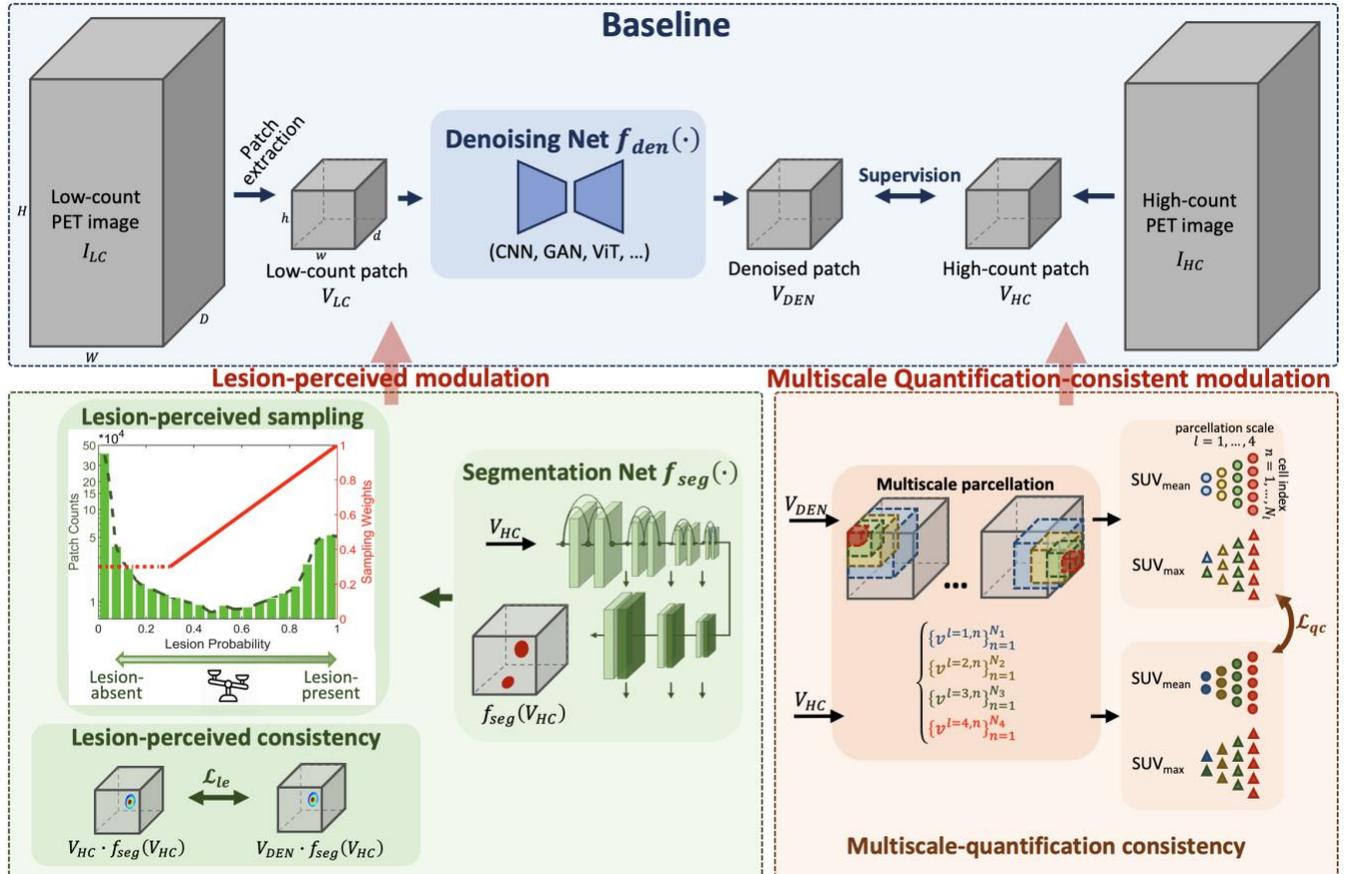

Fig. 1 Illustration of the proposed LeqMod strategy. The lesion-perceived modulation (LeMod) and the multiscale quantification-consistent modulation (QuMod) tune the sampling and optimization process of the denoising baseline during training.

In summary, patches with a higher likelihood of containing lesions alongside heavier background noise are deemed more critical and receive priority in the denoising learning process. Conversely, patches without lesions and less noise are selected less frequently, adjusting the denoising learning process to focus more on maintaining or recovering lesion information.

*2) Lesion-perceived consistency constraint*

Beyond the imbalance between the amounts of patches with lesions and those without, there also exists a disparity between lesional and non-lesional voxels. Often, in a patch identified as containing lesions, the majority of voxels are non-lesional. To further emphasize lesion recovery during the denoising process, the LeMod introduces a lesion-perceived consistency constraint, $\mathcal{L}_{le}$, which provides additional rigorous supervision specifically over lesion voxels, as shown in Fig. 1. $\mathcal{L}_{le}$ is formulated as:

$$\mathcal{L}_{le}(V_{DEN}, V_{HC}) = \frac{1}{\mathcal{J}} \sum_{j=1}^{\mathcal{J}} f_{seg}(V_{HC}(j)) \cdot |V_{DEN}(j) - V_{HC}(j)| \quad (2)$$

where $\mathcal{J}$ denotes the total number of lesion voxels determined by the segmentation network $f_{seg}(\cdot)$. $\mathcal{L}_{le}$ concentrates the optimization specifically on lesion areas and prioritizes voxels based on their probability of being lesional. It gives higher attention to voxels that are more likely to be part of lesions. Such a design ensures that within lesions, the denoised output $V_{DEN}$ aligns closely with the high-count reference $V_{HC}$ voxel-by-voxel in terms of SUV. The emphasis on voxel-wise lesion SUV consistency also guarantees the accuracy of the total lesion glycolysis (TLG) metric [32], which is the multiplication of tumor volume and average SUV in tumor. Keeping the TLG accuracy during denoising is of great clinical importance.

*B. Multiscale Quantification-Consistent Modulation*

The second component of the LeqMod, the QuMod, aims at preserving radioactivity quantification accuracy on any selected sub-regions across the entire image scope, thus improving overall image quality and mitigating over-smoothing. The QuMod formulates this expectation into a loss function to guide denoising learning.

As illustrated in Fig. 1, the QuMod performs a multiscale parcellation to mimic semantic division, where sub-volumes of varying sizes slide across the PET image with overlaps, representing the image as a collection of numerous volumetric sub-regions at different scales. Following this parcellation, a multiscale-quantification consistency constraint, $\mathcal{L}_{qu}$, is conducted on these sub-volumes. The $\mathcal{L}_{qu}$ demands that across multiple parcellation scales, each sub-volume derived from the denoised output $V_{DEN}$ matches in SUV$_{mean}$ and SUV$_{max}$ with its counterpart from the high-count reference $V_{HC}$, expressed as:

$$\mathcal{L}_{qu} = \sum_{l=1}^{4} \frac{\mu_l}{N^l} \cdot \sum_{n=1}^{N^l} \left( \left| \frac{1}{\mathbb{J}^{l,n}} \sum_{j=1}^{\mathbb{J}^{l,n}} v_{DEN}^{l,n}(j) - \frac{1}{\mathbb{J}^{l,n}} \sum_{j=1}^{\mathbb{J}^{l,n}} v_{HC}^{l,n}(j) \right| + \left| \max_j v_{DEN}^{l,n}(j) - \max_j v_{HC}^{l,n}(j) \right| \right) \quad (3)$$

where $l = 1,2,3,4$ denotes the parcellation scale indices. $N^l$ represents the total number of sub-volumes at the $l$-th scale, with $n$ as the index for individual sub-volume. $\{v_{DEN}^{l,n}\}$ and $\{v_{HC}^{l,n}\}$ are sub-volumes obtained from $V_{DEN}$ and $V_{HC}$, respectively. $\mathbb{J}^{l,n}$ indicates the total voxel count within $v_{HC}^{l,n}$, with $j$ as the voxel index. The sub-volume size for parcellation at scales $l = 1,2,3,4$ is determined as fractions $(\frac{1}{2}, \frac{1}{4}, \frac{1}{8}, \frac{1}{16})$ of the patch dimension $h \times w \times d$. The sliding stride is configured to be half of the corresponding sub-volume size. The parameter $\mu_l$ denotes weights allocated for each scale. Considering that constraints on smaller sub-volumes also beneficially affect larger sub-volumes, $\mu_l$ is set to (0.03, 0.07, 0.15, 0.75) for scales $l = 1,2,3,4$, with the highest weight on finest parcellation [33][34].

*C. Model-Adaptable Optimization*

The LeqMod can be flexibly applied to various denoising baseline models. The overall optimization process for models utilizing CNNs and ViTs follows Eq. (4). For models based on GANs, the optimization adheres to Eq. (5).

$$\min \mathcal{L}: \mathcal{L} = \mathcal{L}_{base} + \lambda_{le}\mathcal{L}_{le} + \lambda_{qu}\mathcal{L}_{qu}$$
$$\mathcal{L}_{base} = \frac{1}{\mathcal{N}} \sum_{j=1}^{\mathcal{N}} \|V_{DEN}(j) - V_{HC}(j)\|_2 \quad (4)$$

$$\min_G \max_D (\mathcal{L} + \mathcal{L}_{GAN})$$
$$\mathcal{L}_{GAN} = \mathbb{E}\left[\log\left(1 - D(G(V_{LC}))\right)\right] + \mathbb{E}[\log D(V_{HC})] \quad (5)$$

$\mathcal{L}_{base}$ adopts the L2 loss between the denoised output $V_{DEN}$ and the high-count $V_{HC}$. $\mathcal{N}$ denotes the total voxel count in $V_{DEN}$, with $j$ as the voxel index. $\lambda_{le}$ and $\lambda_{qu}$ are scalar weights for corresponding terms $\mathcal{L}_{le}$ and $\mathcal{L}_{qu}$, respectively. In the context of GANs, $G$ and $D$ refer to the generator and the discriminator, respectively. $\mathcal{L}_{GAN}$ is the adversarial loss [35] for gaming between $G$ and $D$.

III. EXPERIMENTS

*A. Materials*

The proposed LeqMod was evaluated on large datasets from multiple medical centers and vendors. Datasets A, B, and C were specifically collected for the denoising task, including pairs of low- and high-count images, where low-count PET data at various count levels were created by non-overlapping down-sampling the PET list-mode data. **Dataset A** was acquired at the Department of Nuclear Medicine, University of Bern, Switzerland [13], including 209 $^{18}$F-FDG subjects scanned on a

5TABLE I Details of used datasets.

| | For main denoising | | | For auxiliary segmentation |
|---|---|---|---|---|
| | **Dataset A** | **Dataset B** | **Dataset C** | **Dataset D** |
| Medical center | University of Bern, Switzerland | Ruijin Hospital, China | Yale-New Haven Hospital, USA | University Hospital Tübingen, Germany |
| Vendor | Siemens Biograph Vision Quadra | United Imaging uExplorer | Siemens Biograph mCT | Siemens Biograph mCT |
| Average dose (mean±SD) | 264.1±18.2 MBq | 260.1±12.2 MBq | 256.3±16.2 MBq | 314.7±22.1 MBq |
| Tracer | $^{18}$F-FDG | $^{18}$F-FDG | $^{18}$F-FDG | $^{18}$F-FDG |
| Post-injection uptake time (mean±SD) | 69.8±10.7 min | 64.9±19.8 min | Approximately 60 min | Approximately 60 min |
| Scan duration time | 10 min | 5 min | 5 min / bed position | 2 min / bed position |
| OSEM[1] parameters | 4 iterations, 5 subsets | 4 iterations, 20 subsets | 2 iterations, 21 subsets | 2 iterations, 21 subsets |
| FWHM[2] in Gaussian post-smoothing | 2 mm | No Gaussian filter applied | 5 mm | 2 mm |
| Image size (voxels) | 440×440×644 | 360×360×674 | 440×440×$\hbar$[3] | 440×440×$\hbar$[3] |
| Voxel size (mm$^3$/voxel) | 1.65×1.65×1.65 | 1.67×1.67×2.89 | 2.04×2.04×2.03 | 2.04×2.04×3 |
| Low-count levels (%) | 1,2,5,10,25,50 | 1,2,5,10,25,50 | 5,10,20 | / |
| # cases | 209 | 320 | 200 | 501 |
| Train/Val/Test | 100/10/99 | 150/15/155 | 0/0/200 | 480/21/0 |

[1] OSEM: ordered-subsets expectation maximization algorithm.
[2] FWHM: full width at half maximum.
[3] $\hbar$ depends on patient height.

Siemens Biograph Vision Quadra scanner. **Dataset B** was collected at Ruijin Hospital, Shanghai, China [13], including 320 $^{18}$F-FDG subjects scanned on a United Imaging uExplorer scanner. **Dataset C** was obtained from the Yale-New Haven Hospital, USA [17], including 200 $^{18}$F-FDG subjects scanned on a Siemens Biograph mCT scanner. **Dataset D** is a publicly available PET lesion segmentation dataset [36]. It was used for pre-training the auxiliary lesion segmentation network in this study. This dataset was acquired at University Hospital Tübingen, Germany, using a Siemens Biograph mCT scanner. It included 501 patient studies, each featuring at least one FDG-avid tumor lesion. An experienced radiologist and a nuclear medicine physician gave manual annotations, resulting in image-paired segmentation masks.

Details of datasets are listed in Table I.

### B. Technical Details

All PET images underwent cropping pre-processes to remove the outermost background areas lacking informative content, minimizing effects of voxels outside the body. For network processing, the entire image was divided into patches using a patch size of 80×80×80 voxels and a stride size of 20×20×20 voxels. Training, validation, and testing were conducted on mixed data from multiple medical centers, with the dataset split detailed in Table I. The auxiliary lesion segmentation network $f_{seg}(\cdot)$ was pre-trained using Dataset D with rich lesion information and 200 extra lesion-absent patches cropped from training sets of Dataset A and B. These extra patches were selectively extracted from regions surrounding the brain, bladder, and heart which are prone to be mistakenly identified as lesions by DL models [36]. Including these patches into segmentation training helps the network $f_{seg}(\cdot)$ reduce false positives and improve its generalizability to datasets from different sources. Note that the $f_{seg}(\cdot)$ along with its generated lesion labels are used solely during the training phase of the denoising model and are not required for denoising inference stage.

The LeqMod was deployed on three distinct denoising frameworks covering CNN, ViT, and GAN. Specifically, Unet [30] and SwinUNETR [37] were selected as representatives for CNN and ViT, respectively. In the GAN configuration, the generator and discriminator architectures adopted Unet and patchGAN [35], respectively. All models were implemented on PyTorch platform using two Nvidia A40 GPUs. The Adam optimizer with an initial learning rate of $10^{-4}$ was adopted. Learning rate was reduced by a factor of 0.1 if no decrease in loss was observed over five consecutive epochs. Training was terminated when the learning rate fell below $10^{-7}$.

### C. Evaluation Measures

We use three types of metrics to assess the PET image denoising performance.

The first type includes widely used image quality metrics [38]: the normalized root-mean-square error (NRMSE), peak signal-to-noise ratio (PSNR), and structural similarity index (SSIM). These metrics are employed to evaluate the overall quality of the denoised image $I_{DEN}$, in comparison to its corresponding high-count reference $I_{HC}$. Lower NRMSE values and higher PSNR and SSIM scores are preferred, indicating a closer resemblance to the high-count images and a higher level of image fidelity.

The second type evaluates quantification biases in clinical metrics between $I_{DEN}$ and $I_{HC}$, including SUV$_{mean}$ and SUV$_{max}$ for individual lesions, as well as whole-body TLG for each subject. Lesion labels used in calculations for Dataset A and B were determined by the trained segmentation network $f_{seg}(\cdot)$ on $I_{HC}$, compensating for the absence of manual annotations. Further, for assessing performance on all lesions, including those might be mistaken by segmentation network, lesion labels for Dataset C were manually annotated by two physicians working collaboratively. Zero-approaching biases in lesion SUV$_{mean}$, SUV$_{max}$, and TLG are desirable, indicating higher credibility of denoised images for clinical usage.

The third type evaluates the lesion contrast and visibility in denoised image $I_{DEN}$, with the trained segmentation network $f_{seg}(\cdot)$ serving as an observer. This evaluation approach is inspired by similar settings in [39][40] that evaluate DL-generated images using downstream segmentation. Applying $f_{seg}(\cdot)$ to both $I_{DEN}$ and the reference $I_{HC}$, the Tversky index is computed between segmentations:

$$Tversky(S_{DEN}, S_{HC}) = \frac{|S_{DEN} \cap S_{HC}|}{|S_{DEN} \cap S_{HC}| + \alpha|S_{DEN} \setminus S_{HC}| + \beta|S_{HC} \setminus S_{DEN}|} \quad (6)$$

where $S_{DEN} = B(f_{seg}(I_{DEN}))$ and $S_{HC} = B(f_{seg}(I_{HC}))$ denote segmentations performed on $I_{DEN}$ and $I_{HC}$, respectively. $B(\cdot)$ represents the binary thresholding by 0.5, obtaining binary segmentation masks from the lesion probability maps output by $f_{seg}(\cdot)$. $\alpha$ and $\beta$ specify the penalties for false positives and false negatives, respectively. Tversky index with $\alpha = \beta = 0.5$ equals to the Dice coefficient. A higher Tversky index is expected, indicating better lesion visibility in denoised images.

### D. Experimental setup

In Section IV-A, we demonstrate the effectiveness of the LeqMod in enhancing both image quality and quantification accuracy. Section IV-B presents evidence of improved lesion visibility attributed to the LeqMod. In Section IV-C, detailed ablation studies are conducted to illustrate individual contributions of each LeqMod component. Section IV-D validates the versatile applicability of the LeqMod across different baseline models.

## IV. RESULTS

### A. LeqMod Effectiveness

A grid search across a range of values was conducted on validation to determine the optimal hyperparameters $\lambda_{le}$ and $\lambda_{qu}$ in LeqMod, with results visualized in Fig. 2, using Unet as the sample denoising baseline. As shown, across a broad range of hyperparameter selections, incorporating LeqMod consistently outperforms the baseline without LeqMod, with the setting of $\lambda_{le} / \lambda_{qu} = 0.15/0.5$ yielding the best overall performance. Unless otherwise specified, all results presented in the following sections are based on this hyperparameter setting.

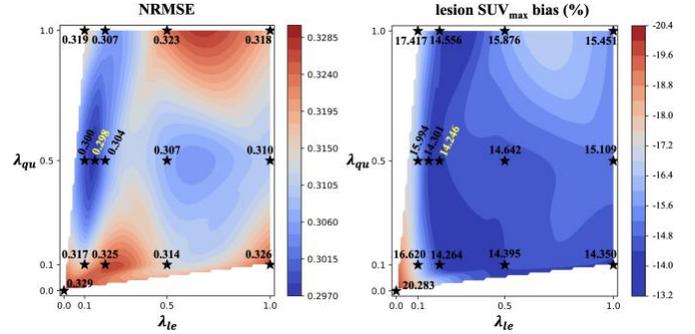

Fig. 2. Performance of LeqMod in relation to hyperparameters $\lambda_{le}$ and $\lambda_{qu}$. Lower values (indicated by bluer colors) represent better performance.

The comprehensive test performance of LeqMod applied to Unet is detailed in Table II. Results illustrate that the LeqMod significantly improves nearly all the metrics, across a variety of scanners and noise levels. This improvement indicates contributions of the LeqMod to enhanced visual similarity with high-count references and more accurate lesion quantification in denoised images. Particularly, for denoising extremely noisy images at 1% low-count levels, the LeqMod reduces quantification biases in lesion $SUV_{mean}$, $SUV_{max}$, and TLG by 5.29%, 10.44%, and 2.56% on average, respectively.

TABLE II Effectiveness of the LeqMod in PET image denoising, across multiple vendors and varying levels of noise. Metrics include NRMSE, PSNR, SSIM, lesion $SUV_{mean}$ bias, $SUV_{max}$ bias, and TLG bias.

| Scanner | Quadra | | | | | | uExplorer | | | | | | mCT | | |
|---|---|---|---|---|---|---|---|---|---|---|---|---|---|---|---|
| count level | 1% | 2% | 5% | 10% | 25% | 50% | 1% | 2% | 5% | 10% | 25% | 50% | 5% | 10% | 20% |
| NRMSE (↓) | | | | | | | | | | | | | | | |
| low-count | *0.848* | *0.562* | *0.351* | *0.253* | *0.159* | *0.096* | *1.863* | *1.064* | *0.621* | *0.441* | *0.266* | *0.160* | *1.151* | *0.784* | *0.521* |
| w/o | 0.278 | 0.237 | 0.211 | 0.193 | 0.130 | 0.096 | 0.568 | 0.464 | 0.404 | 0.296 | 0.220 | 0.162 | 0.519 | 0.438 | 0.350 |
| w | **0.267**† | **0.223**† | **0.196**† | **0.174**† | **0.125**† | **0.088**† | **0.541**† | **0.436**† | **0.387**† | **0.286**† | **0.209**† | **0.147**† | **0.510**† | **0.427**† | **0.342**† |
| PSNR (↑) | | | | | | | | | | | | | | | |
| low-count | *53.496* | *53.491* | *53.145* | *54.601* | *57.767* | *61.920* | *42.709* | *44.554* | *47.718* | *50.062* | *53.933* | *58.116* | *44.487* | *46.325* | *48.944* |
| w/o | 52.405 | 53.883 | 55.424 | 56.640 | 58.903 | 61.413 | 46.154 | 48.224 | 50.352 | 52.048 | 54.431 | 57.937 | **47.306** | 48.809 | 50.731 |
| w | **52.762**† | **54.051**† | **55.743**† | **56.850**† | **59.190**† | **62.220**† | **46.835**† | **48.656**† | **50.745**† | **52.285**† | **54.968**† | **58.038**† | 47.285 | **48.899**† | **50.857**† |
| SSIM (↑) | | | | | | | | | | | | | | | |
| low-count | *92.289* | *94.543* | *97.934* | *99.152* | *99.774* | *99.919* | *89.309* | *94.320* | *97.497* | *98.655* | *99.479* | *99.808* | *92.676* | *95.641* | *97.759* |
| w/o | 99.445 | 99.413 | **99.479** | 99.631 | 99.808 | 99.896 | 97.458 | 97.702 | 99.050 | 99.251 | 99.590 | 99.726 | 97.952 | 98.415 | 98.894 |
| w | **99.450**† | **99.639**† | 99.429 | **99.769**† | **99.867**† | **99.931**† | **98.016**† | **98.750**† | **99.121**† | **99.440**† | **99.674**† | **99.822**† | **98.091**† | **98.588**† | **99.003**† |
| SUV$_{mean}$ bias (%) (o) | | | | | | | | | | | | | | | |
| low-count | *-2.882* | *-0.444* | *1.189* | *1.703* | *1.488* | *1.052* | *-4.754* | *-0.780* | *1.180* | *1.378* | *0.917* | *0.815* | *-3.953* | *-2.137* | *-0.802* |
| w/o | -20.948 | -15.127 | -9.445 | -6.694 | -3.858 | -3.209 | -34.051 | -25.021 | -15.913 | -11.023 | -7.318 | -5.933 | -25.517 | -20.095 | -13.199 |
| w | **-17.128**† | **-12.840**† | **-8.491**† | **-5.711**† | **-3.131**† | **-1.557**† | **-28.726**† | **-20.591**† | **-12.737**† | **-8.471**† | **-4.192**† | **-2.494**† | **-23.469**† | **-17.493**† | **-11.657**† |
| SUV$_{max}$ bias (%) (o) | | | | | | | | | | | | | | | |
| low-count | *57.743* | *30.397* | *16.241* | *10.933* | *6.205* | *3.271* | *88.944* | *38.594* | *17.863* | *10.998* | *5.399* | *2.929* | *24.004* | *13.675* | *6.638* |
| w/o | -26.473 | -20.177 | -13.468 | -9.816 | -6.474 | -6.318 | -43.518 | -32.297 | -20.356 | -14.586 | -10.081 | -8.942 | -28.522 | -23.243 | -16.480 |
| w | **-19.274**† | **-15.108**† | **-10.181**† | **-6.959**† | **-3.809**† | **-2.546**† | **-31.132**† | **-23.759**† | **-15.350**† | **-10.463**† | **-4.848**† | **-2.985**† | **-24.676**† | **-18.803**† | **-13.296**† |
| TLG bias (%) (o) | | | | | | | | | | | | | | | |
| low-count | *-1.963* | *-0.742* | *0.174* | *0.545* | *0.596* | *0.469* | *-2.362* | *0.127* | *1.382* | *1.368* | *0.921* | *0.769* | *-1.154* | *-0.476* | *-0.360* |
| w/o | -5.142 | -3.772 | -2.446 | -2.076 | -1.311 | -1.055 | -13.643 | -8.975 | -5.463 | -3.591 | -2.495 | -2.081 | -8.386 | -6.179 | -4.172 |
| w | **-3.983**† | **-3.261**† | **-2.394**† | **-1.671**† | **-1.281**† | **-0.692**† | **-10.058**† | **-6.332**† | **-3.487**† | **-2.321**† | **-1.251**† | **-0.744**† | **-7.415**† | **-5.169**† | **-3.557**† |

† $P<0.05$ based on the Wilcoxon signed-rank test between groups without (w/o) and with (w) the LeqMod.
(↑), (↓) and (o) represent expecting higher, lower, and zero-approaching values, respectively.

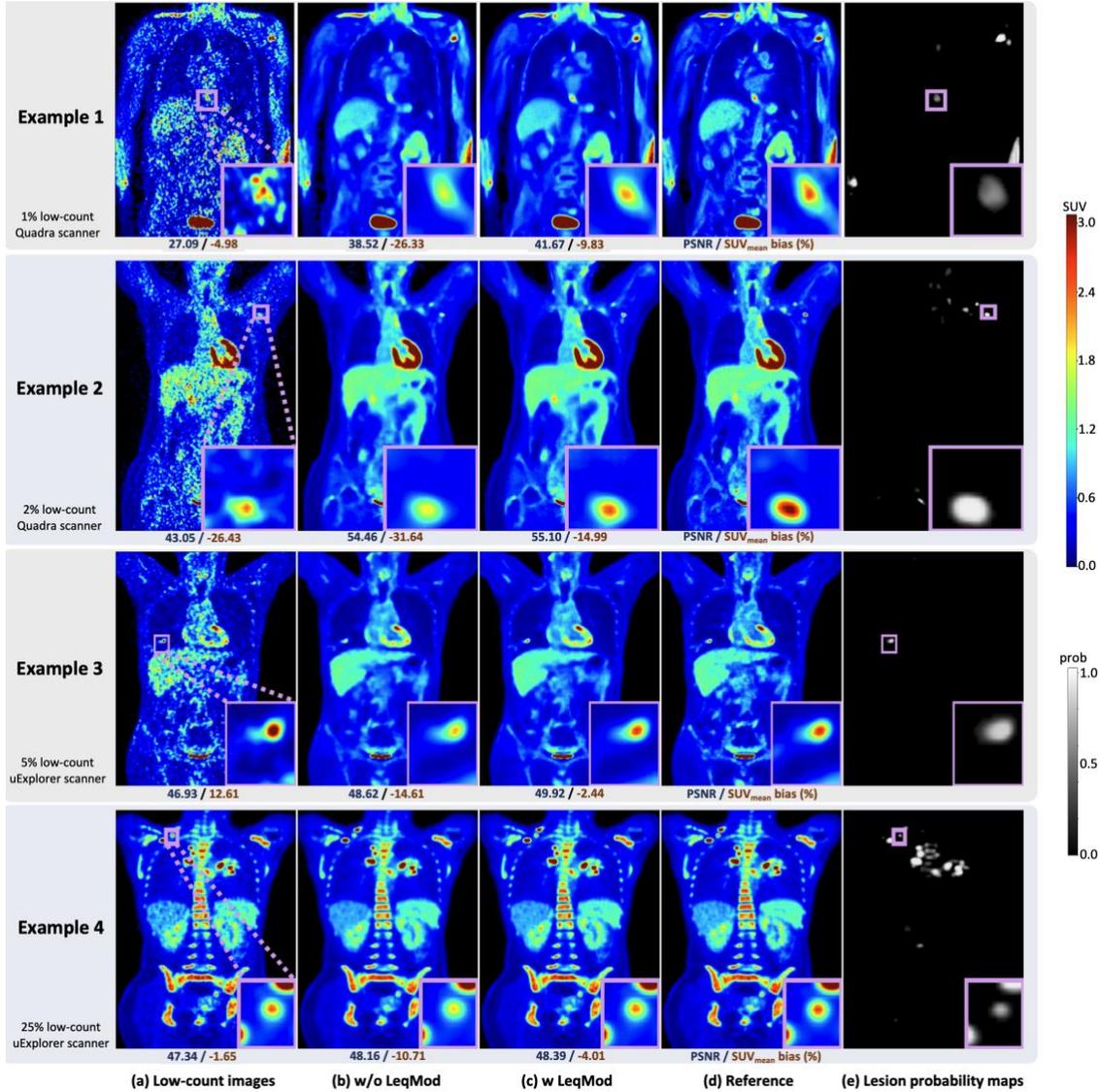

Fig. 3 Visual comparisons of denoised results without (w/o) and with (w) the LeqMod. Lesion probability maps determined by the auxiliary segmentation network are also displayed for reference. Sub-regions including lesions or lesion-like hot spots are cropped, amplified, and displayed in the bottom-right corner of each image. PSNR and lesion $SUV_{mean}$ bias values, compared to the high-count reference, are denoted below each image.

Fig. 3 presents sample denoised images without and with the LeqMod. Without the LeqMod, denoised images frequently exhibit over-smoothing, leading to detail loss and poor contrast in small lesions and areas resembling lesions. Incorporation of the LeqMod mitigates this over-smoothing effect and enhances the contrast of lesions or lesion-like hot spots in denoised images, as shown in Fig. 3.

### B. Improved Lesion Visibility

The LeqMod leads to improved lesion visibility in denoised images, taking all test cases of the 5% low-count level as examples, as evidenced in Table III. Using the trained lesion segmentation network as an observer, the segmentation on images of 5% low-count level significantly deviates from that on high-count references, with notably low Tversky indexes of 0.40, 0.34, and 0.30 for coefficients $\alpha/\beta = 0.3/0.7$, $\alpha/\beta = 0.5/0.5$, and $\alpha/\beta = 0.7/0.3$, respectively. This indicates that noise-contaminated low-count images suffer from poor lesion visibility, limiting their clinical applicability. Denoising low-count images with the Unet baseline improves corresponding Tversky indexes to 0.64, 0.65, and 0.67, demonstrating the efficacy of DL denoising in boosting lesion visibility. The deployment of the LeqMod further enhances this improvement, elevating Tversky indexes to 0.65, 0.66, and 0.69, thereby reinforcing lesion visibility.

TABLE III Improved lesion visibility in denoised images facilitated by the LeqMod. Metrics were averaged across all test cases from the three scanners at the 5% low-count level.

|  | Tversky Index (↑) | | |
| --- | --- | --- | --- |
|  | $\alpha=0.3, \beta=0.7$ | $\alpha=0.5, \beta=0.5$ (Dice) | $\alpha=0.7, \beta=0.3$ |
| low-count | 0.395±0.274 | 0.338±0.252 | 0.301±0.237 |
| w/o LeqMod | 0.637±0.194 | 0.653±0.185 | 0.674±0.175 |
| w LeqMod | **0.646±0.195**† | **0.664±0.182**† | **0.688±0.171**† |

† $P<0.05$ based on the Wilcoxon signed-rank test between the w/o and w LeqMod groups.
(↑) represents expecting higher values.

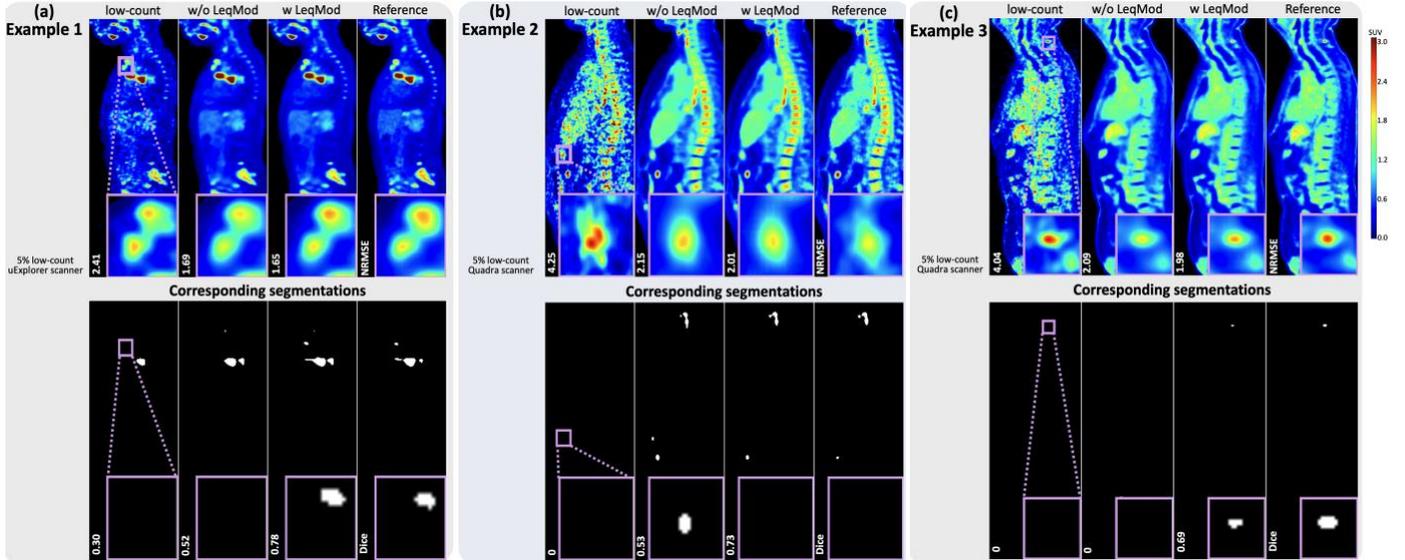

Fig. 4 Comparisons of lesion segmentation on denoised images without (w/o) and with (w) the LeqMod. Lesion probability maps output by the segmentation network were binarized by 0.5 to obtain segmentation masks. Sub-regions of interest are cropped, amplified, and displayed in the bottom-right corner of each image. NRMSE values and Dice coefficients are denoted in white text.

Fig. 4 presents examples of lesion segmentation on low-count and denoised images. Observations from the segmentation network reveal that denoised images processed with LeqMod exhibit the greatest similarity to high-count references, ensuring more accurate downstream lesion segmentations. The LeqMod reduces both lesion false negatives (Fig. 4 (a)(c)) and false positives (Fig. 4 (b)) in PET image denoising.

### C. Ablation Studies

To assess the contribution of each component within the LeqMod, ablation studies were conducted, and the average results across all test cases are listed in Table IV.

TABLE IV Ablation studies on the LeqMod using Unet as a baseline. Metrics were averaged across all test cases from the three scanners.

|  | NRMSE (↓) | PSNR (↑) | SSIM (↑) | $SUV_{mean}$ bias (%) (o) | $SUV_{max}$ bias (%) (o) | TLG bias (%) (o) |
|---|---|---|---|---|---|---|
| Baseline | 0.339 ±0.413 | 52.197 ±7.094 | 98.941 ±2.095 | -13.782 ±18.877 | -18.105 ±22.735 | -5.062 ±7.679 |
| +LeMod | 0.331 ±0.278 | 52.388 ±7.067 | 99.040 ±1.879 | -12.802 ±17.650 | -14.244 ±22.251 | -4.072 ±6.079 |
| +QuMod | 0.322 ±0.634 | 52.510 ±7.137 | 99.103 ±1.454 | -13.110 ±18.497 | -16.779 ±21.947 | -4.433 ±7.723 |
| +LeqMod | **0.314** ±0.279 | **52.555** ±7.156 | **99.135** ±1.378 | **-10.546** ±15.618 | **-12.190** ±19.778 | **-3.725** ±6.325 |

(↑), (↓) and (o) represent expecting higher, lower, and zero-approaching values, respectively.

As shown in Table IV, LeMod and QuMod exhibit complementary contributions to lesion quantification accuracy and the overall fidelity of denoised images. LeMod, with its specific focus on lesions, reduces the lesion $SUV_{max}$ bias by 3.86% on average. However, due to the small proportion of lesions relative to the entire image, its impact on global image quality metrics such as NRMSE, PSNR, and SSIM is limited. On the other hand, QuMod emphasizes volume-by-volume radioactivity consistency across the entire image, leading to notable improvements in NRMSE, PSNR, and SSIM, though its influence on lesion quantification biases is comparatively modest. This observation is further supported by Fig. 2, which demonstrates that the parameter $\lambda_{le}$ for LeMod has a more pronounced impact on reducing lesion SUV bias, while $\lambda_{qu}$ for QuMod primarily influences the overall NRMSE.

By integrating both components, LeqMod achieves the best performance, yielding a mean reduction of 5.92% in lesion $SUV_{max}$ bias and a mean increase of 0.36 in PSNR, surpassing the baseline in both lesion quantification accuracy and overall image fidelity.

Fig. 5 is the Bland-Altman analysis on the performance of the LeqMod and its ablated versions, assessing the agreement between denoised images and high-count references in calculating TLG. Results show that both the LeMod and QuMod contribute to a higher agreement between denoised images and references. The LeqMod achieves the highest agreement, outperforming the baseline with a reduction of 0.11 in the average differences.

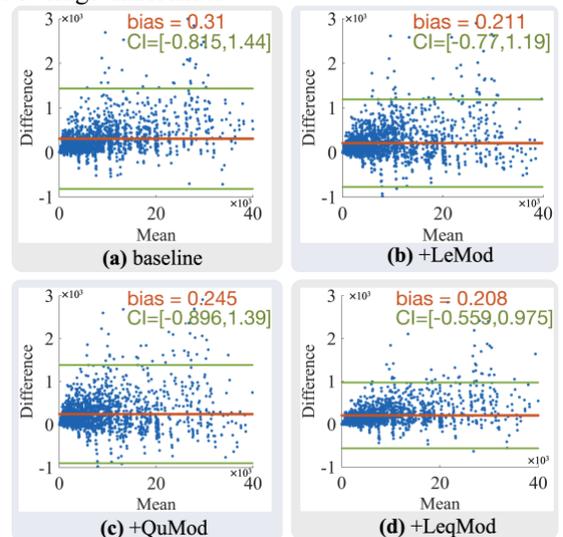

Fig. 5 Bland-Altman analysis was conducted on all test cases from three scanners to evaluate the performance of LeqMod and its ablated versions in quantifying TLG. The y-axis represents the bias of TLG values, while the x-axis represents the average TLG values measured by denoised images and high-count references. The mean bias is highlighted with a red line, and the 96% confidence intervals (CI) of the bias are indicated with green lines.

### D. Adaptability to Various Architectures

Beyond its effectiveness on the Unet baseline, the LeqMod shows flexible adaptability to a variety of denoising frameworks. Table V summarizes the performance of the LeqMod on SwinUNETR and GAN architectures, indicating enhancements in both image quality and lesion quantification metrics. Averaged across all test cases from the three scanners, the LeqMod facilitates a TLG bias reduction of 0.93% for the GAN baseline and 1.34% for the SwinUNETR baseline.

TABLE V Adaptability of the LeqMod to different denoising baselines.

| | | GAN | | | | SwinUNETR | | | |
|---|---|---|---|---|---|---|---|---|---|
| | | NRMSE (↓) | | TLG bias (%) (o) | | NRMSE (↓) | | TLG bias (%) (o) | |
| | | w/o | w | w/o | w | w/o | w | w/o | w |
| Quadra | 1% | 0.273 | **0.266**† | -5.115 | **-4.444**† | 0.294 | **0.284**† | -7.395 | **-5.304**† |
| | 2% | 0.252 | **0.234**† | -4.092 | **-3.376**† | 0.240 | **0.231**† | -5.617 | **-3.889**† |
| | 5% | 0.196 | **0.194** | -2.592 | **-2.328**† | **0.207** | 0.216 | -4.244 | **-2.980**† |
| | 10% | 0.178 | **0.169**† | -1.787 | **-1.583**† | 0.184 | **0.180** | -3.312 | **-2.133**† |
| | 25% | 0.146 | **0.131**† | -1.035 | **-0.879**† | 0.173 | **0.137**† | -2.193 | **-1.163**† |
| | 50% | 0.100 | **0.094** | -0.699 | **-0.610** | 0.140 | **0.114**† | -2.095 | **-0.964**† |
| uExplorer | 1% | 0.551 | **0.536**† | -13.231 | **-9.671**† | 0.574 | **0.555**† | -15.354 | **-13.586**† |
| | 2% | 0.439 | **0.432**† | -8.601 | **-6.696**† | 0.453 | **0.440**† | -11.286 | **-8.921**† |
| | 5% | 0.354 | **0.353**† | -4.869 | **-4.163**† | 0.371 | **0.349**† | -7.069 | **-5.080**† |
| | 10% | 0.284 | **0.274**† | -3.285 | **-3.092**† | 0.303 | **0.294**† | -5.371 | **-3.672**† |
| | 25% | **0.213** | 0.224 | -1.800 | **-1.679**† | 0.231 | **0.226**† | -3.721 | **-2.249**† |
| | 50% | 0.153 | **0.142**† | **-1.016** | -1.028 | 0.182 | **0.165**† | -2.805 | **-1.351**† |
| mCT | 5% | 0.538 | **0.527**† | -8.466 | **-7.428**† | 0.531 | **0.524**† | -8.485 | **-7.507**† |
| | 10% | 0.523 | **0.449**† | -5.828 | **-5.056**† | 0.459 | **0.430**† | -6.070 | **-4.990**† |
| | 20% | 0.344 | **0.339**† | -3.604 | **-3.037** | 0.361 | **0.348**† | -4.187 | **-3.712** |

† $P<0.05$ based on the Wilcoxon signed-rank test between the w/o and w LeqMod groups.
(↓) and (o) represent expecting lower and zero-approaching values, respectively.

For more intuitive comparisons, Fig. 6 employs boxplots to depict lesion quantification accuracy across different architectures with and without the LeqMod. In every scenario, setups incorporating the LeqMod outperform those without, with the GAN enhanced by LeqMod leading in overall performance. Further, Fig. 7 presents a visual comparison of denoising outcomes from various models, clearly demonstrating that the LeqMod improves denoising performance across multiple baseline models.

## V. DISCUSSION

The LeqMod introduces innovative ideas for developing clinically reliable and practically viable DL models for denoising PET images.

From a clinical perspective, the paramount attributes of denoised images are lesion visibility, quantification accuracy, and the absence of hallucinations such as false positive lesions. Previous DL denoising models [10]-[25] for PET have focused on producing visually striking images, thereby enhancing visual image quality metrics like PSNR and SSIM. They often overlook the clinical utility of denoised images for subsequent lesion analysis and diagnosis. While these methods yield high-resolution images, the potential loss of lesions, biased radioactivity quantification, and hallucination generations in denoised outputs compromise the DL credibility for clinical uses. The LeqMod pioneered in integrating lesion segmentation and multiscale quantification to assist PET image denoising, with an emphasis on factors essential for clinical evaluation. Consequently, it improves lesion contrast and visibility in denoised images and elevates quantification accuracy for clinical metrics such as $SUV_{mean}$, $SUV_{max}$, and TLG, as evidenced in Table II and III and Fig. 3 and 4. Compared to the baseline, integrating the LeqMod results in a 5.92% decrease in lesion $SUV_{max}$ bias, averaged across all test cases from the three scanners. Since these metrics are significantly associated with overall survival [41], and even slight variability in SUV measurements might cause misinterpretation of tumor progression and therapy response [42], reducing biases is crucial for potential better clinical decision-making.

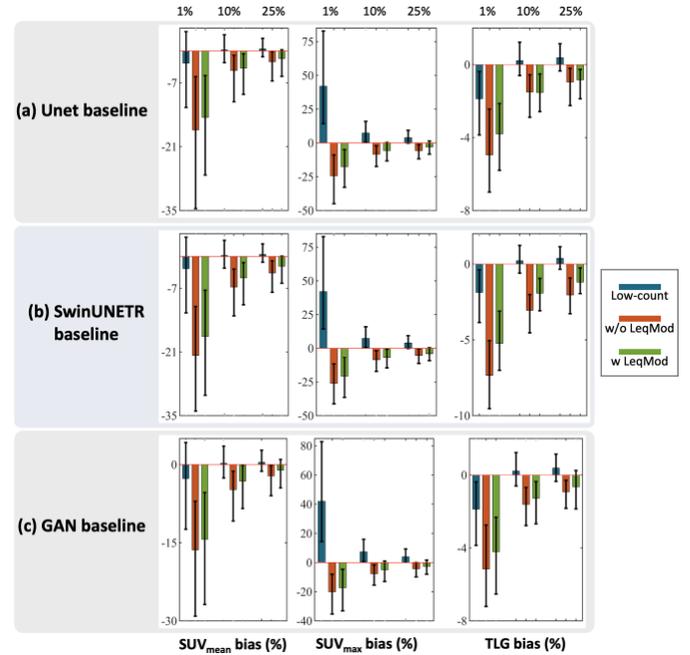

Fig. 6 Boxplots on LeqMod performance across various denoising architecture baselines: **(a)** Unet, **(b)** SwinUNETR, and **(c)** GAN, using test cases from the Quadra scanner as examples. Metrics are lesion $SUV_{mean}$ bias, $SUV_{max}$ bias, and TLG bias.

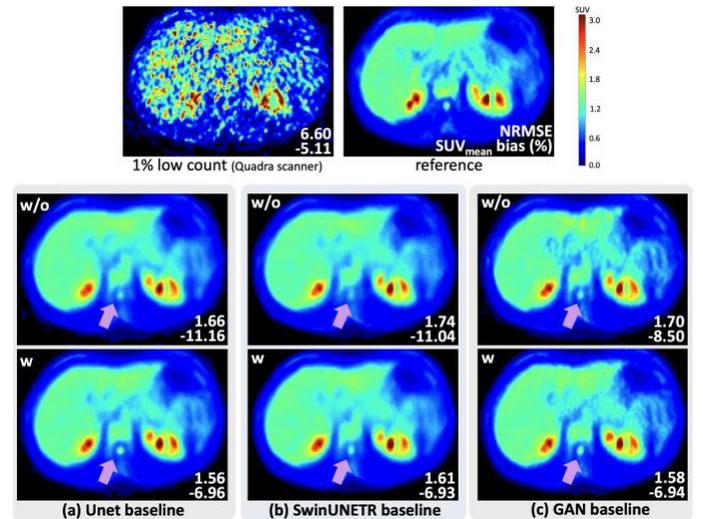

Fig. 7 Adaptability of the LeqMod strategy to different denoising architectures.

From a practical perspective, earlier segmentation-assisted denoising models [26]-[28] have typically integrated semantic contexts by adding extra convolutional layers into denoising architectures, which leads to a marked increase in computational complexity. They require manually preparing segmentation labels matched to noisy-clean image pairs in denoising datasets. In contrast, the LeqMod enhances denoising performance without adding computational burden to model inference, by modulating data distributions and optimization processes during model training. It decouples the auxiliary lesion segmentation from the main denoising architecture and leverages public data sources to pre-train segmentation network, thereby avoiding manual lesion label requirements matched to low- and high-count images in denoising datasets. These designs make the LeqMod a highly adaptable, plug-and-play module that is compatible with various denoising architectures. Shown in Table V and Fig. 6 and 7, the LeqMod improves denoising performance across different baselines, including Unet, SwinUNETR, and GAN.

The LeqMod benefits from and is characterized by its two components, the LeMod and the QuMod.

The LeMod is inspired by active learning principles [43]. Active learning strategically selects the most informative and representative data samples for annotating and model learning, which enhances model performance while minimizing labeling and training expenses. This approach is particularly valuable in addressing imbalances by prioritizing informative but numerically inferior samples. In the context of PET image denoising, lesion-present samples are more representative but less numerous. At the patch level, lesion-absent patches significantly outnumber those containing lesions. At the voxel level, lesion voxels are less common than non-lesion ones. To counteract these imbalances, the LeMod performs active learning in forms of patch-level lesion-perceived sampling and voxel-level lesion-perceived consistency. The former adjusts training data distributions to mitigate patch-level imbalances, selecting patches based on the segmentation network's confidence in lesion presence. The latter modifies loss functions to concentrate on lesional voxels identified by the segmentation network. Importantly, the LeMod utilizes continuous segmentation probabilities instead of binary segmentation masks thresholded at 0.5, allowing for more nuanced modulation on denoising training and reducing the dependency on segmentation accuracy. As a result, the LeMod has been shown to improve lesion quantification accuracy, as detailed in Table IV and Fig. 5. Averaged across all test cases from the three scanners, the LeMod reduces lesion $SUV_{mean}$ bias, $SUV_{max}$ bias, and TLG bias by 0.98%, 3.86%, and 0.99%, respectively.

Beyond focusing on lesions, the QuMod extends to ensure the accuracy of $SUV_{mean}$ and $SUV_{max}$ quantification across multiscale ROIs throughout the entire image. Given the importance of precise radioactivity quantification in anatomical organs and the common lack of image-aligned organ labels, the QuMod employs sub-volumes of varying sizes to mimic anatomical semantics, representing the image as a collection of volumetric ROIs. Then, the QuMod enforces quantification accuracy within each ROI at every scale, with the strictest standards at the finest scale. By imposing constraints across the whole image scope, the QuMod enhances image fidelity and mitigates over-smoothing. Averaged across all test cases from the three scanners, the QuMod improves NRMSE, PSNR, and SSIM by 0.02, 0.31, and 0.16, respectively, and also offers modest reductions in lesion quantification biases, as demonstrated in Table IV and Fig. 5.

Although the LeqMod has shown promising outcomes, there remains substantial space for further advancement. **(1)** The auxiliary segmentation network within LeqMod, while effective, has limitations in its ability to comprehensively and accurately identify lesions. It has been trained on the public dataset [36] of a single scanner featuring tumors of malignant melanoma, lymphoma, and lung cancer. Although it shows a degree of generalizability to other datasets through patch-wise processing, the risk of segmentation inaccuracies (missing certain lesions or mistakenly identifying hot spots as lesions) might weaken the support for denoising. Future versions should aim to encompass wider types of tumors and scanners to provide stronger segmentation-based assistance and explore the impact of segmentation accuracy on denoising performance. Additionally, the impact of segmentation-modulated data distributions (*i.e.*, the balance between lesion and non-lesion information in training sets) on the generation of hallucinated lesion false positives and false negatives warrants further exploration and reliable quantification [44]. **(2)** While LeqMod benefits from decoupling auxiliary segmentation from primary denoising, thereby relaxing dataset requirements, approaches that foster greater dual-task cooperation may offer enhanced mutual benefits, assuming suitable datasets are accessible. Moreover, incorporating anatomical organ semantics from modalities such as CT or MRI, in addition to metabolic lesion knowledge, potentially provides stronger support for denoising. **(3)** LeqMod in this study was applied to a mix of multiple datasets. Future experimental work could explore dataset-specific or domain-adaptive techniques, as well as automatic hyperparameter optimization strategies for improved efficiency. Additionally, with access to datasets containing detailed clinical information, such as cancer staging or survival outcomes, further evaluation could be conducted to investigate the direct impact of reduced lesion SUV bias on downstream clinical decision-making. Such exploration may provide new insights into the method effectiveness and clinical applicability.

## VI. CONCLUSION

We present a lesion-quantification-consistent modulation strategy named the LeqMod, for enhancing low-count PET image denoising with a focus on lesion visibility and precise multiscale radioactivity quantification. We validate the efficacy of each innovation and prove that the LeqMod contributes to improved lesion visibility and reduced SUV biases. The LeqMod strategy, facilitated by downstream analysis, potentially delivers clinical and practical benefits.